\documentclass[usegraphicx,usenatbib,letterpaper]{emulateapj}

\usepackage{ulem}
\usepackage{color}
\usepackage{graphics,graphicx}
\usepackage{times} 
\usepackage{amssymb}
\usepackage{amsmath}
\usepackage{natbib}
\usepackage{url}
\usepackage{multirow}
\usepackage{ctable}
\usepackage{threeparttable}
\newif\ifAMStwofonts
\AMStwofontstrue

\def\xmm{{\it XMM-Newton}}

\def\suzaku{{\it Suzaku}}

\def\chandra{{\it Chandra}}
\def\swift{{\it Swift}}

\def\epicpn{{EPIC-pn}}
\def\epicmos1{{EPIC-MOS1}}
\def\epicmos2{{EPIC-MOS2}}
\def\epicmos{{EPIC-MOS}}

\def\nustar{{\it NuSTAR}}

\def\H0{{\rm ~km~s^{-1}~Mpc^{-1}}}

\def\kev{\hbox{\rm keV}}

\def\ergpcmsqps{\hbox{$\rm\thinspace erg~cm^{-2}~s^{-1}$}}

\def\ergps{\hbox{erg~s$^{-1}$}}

\def\msun{\hbox{$\rm M_{\odot}$}}

\def\chisq{{$\chi^{2}$}}
\def\rchi{{$\chi^{2}_{\nu}$}}

\def\xspec{\hbox{\small XSPEC}}

\def\heasoft{\hbox{\rm{\small HEASOFT}}}
\def\nustardas{\rm {\small NUSTARDAS}}

\def\xmmselect{\hbox{\rm{\small XMMSELECT}}}
\def\ftool{\hbox{\rm{\small FTOOL}}}

\def\addascaspec{\hbox{\rm{\small ADDASCASPEC~\/}}}
\def\sas{\hbox{\rm{\small SAS}}}
\def\epchain{\hbox{\rm{\small EPCHAIN}}}
\def\emchain{\hbox{\rm{\small EMCHAIN}}}

\def\rmfgen{\hbox{\rm{\small RMFGEN}}}
\def\arfgen{\hbox{\rm{\small ARFGEN}}}

\def\addascaspec{\hbox{\rm{\small ADDASCASPEC}}}
\def\nupipeline{\rm{\small NUPIPELINE}}
\def\nuproducts{\rm{\small NUPRODUCTS}}

\def\xrtmkarf{\hbox{\rm {\small XRTMKARF}}}

\def\grid25{\hbox{\rm{\small GRID25}}}
\def\pl{\rm{\small POWERLAW}}

\def\tbabs{\rm{\small TBABS}}
\def\tbnew{\rm{\small TBNEW}}
\def\tbnewlink{http://pulsar.sternwarte.uni-erlangen.de/wilms/research/tbabs}
\def\diskbb{\rm{\small DISKBB}}
\def\diskpbb{\rm{\small DISKPBB}}

\def\cflux{\rm{\small CFLUX}}

\def\etal{et al.}
\def\eg{{\it e.g.~\/}}

\def\ie{{\it i.e.~\/}}

\def\la{\mathrel{\hbox{\rlap{\hbox{\lower4pt\hbox{$\sim$}}}{\raise2pt\hbox{$<$}}}}}
\def\ga{\mathrel{\hbox{\rlap{\hbox{\lower4pt\hbox{$\sim$}}}{\raise2pt\hbox{$>$}}}}}

\def\d25{D$_{25}$}
\def\nh{{$N_{\rm H}$}}

\def\.25{0.25 keV\thinspace}

\def\ordmag{\rm orders of magnitude}

\def\ngc{\rm NGC\,5907 ULX1}
\def\altsrc{{\rm Source S}}
\def\eone{{\rm epoch 1}}
\def\etwo{{\rm epoch 2}}

\shorttitle{\nustar\ and \xmm\ observations of \ngc}
\shortauthors{D.~J. Walton et al.}

\begin{document}

\title{\textit{NuSTAR} and \textit{XMM-Newton} observations of the extreme ultraluminous X-ray source \ngc: A Vanishing Act}

\author{D. J. Walton\altaffilmark{1,2},
F. A. Harrison\altaffilmark{2},
M. Bachetti\altaffilmark{3,4},
D. Barret\altaffilmark{3,4},
S. E. Boggs\altaffilmark{5},
F. E. Christensen\altaffilmark{6},
W. W. Craig\altaffilmark{5},
F. Fuerst\altaffilmark{2},
B. W. Grefenstette\altaffilmark{2},
C. J. Hailey\altaffilmark{7},
K. K. Madsen\altaffilmark{2},
M. J. Middleton\altaffilmark{8},
V. Rana\altaffilmark{2},
T. P. Roberts\altaffilmark{9},
D. Stern\altaffilmark{1},
A. D. Sutton\altaffilmark{9},
N. Webb\altaffilmark{3,4},
W. Zhang\altaffilmark{10},
}
\affil{
$^{1}$ Jet Propulsion Laboratory, California Institute of Technology, Pasadena, CA 91109, USA \\
$^{2}$ Space Radiation Laboratory, California Institute of Technology, Pasadena, CA 91125, USA \\
$^{3}$ Universite de Toulouse; UPS-OMP; IRAP; Toulouse, France \\
$^{4}$ CNRS; IRAP; 9 Av. colonel Roche, BP 44346, F-31028 Toulouse cedex 4, France \\
$^{5}$ Space Sciences Laboratory, University of California, Berkeley, CA 94720, USA \\
$^{6}$ DTU Space, National Space Institute, Technical University of Denmark, Elektrovej 327, DK-2800 Lyngby, Denmark \\
$^{7}$ Columbia Astrophysics Laboratory, Columbia University, New York, NY 10027, USA \\
$^{8}$ Institute of Astronomy, University of Cambridge, Madingley Road, Cambridge CB3 0HA, UK \\
$^{9}$ Department of Physics, Durham University, South Road, Durham DH1 3LE, UK \\
$^{10}$ NASA Goddard Space Flight Center, Greenbelt, MD 20771, USA \\
}

\begin{abstract}
We present results obtained from two broadband X-ray observations of the
extreme ultraluminous X-ray source (ULX) \ngc, known to have a peak X-ray
luminosity of $\sim5\times10^{40}$\,\ergps. These \xmm\ and \nustar\
observations, separated by only $\sim$4 days, revealed an extreme level of
short-term flux variability. In the first epoch, \ngc\ was undetected by
\nustar, and only weakly detected (if at all) with \xmm, while in the second
\ngc\ was clearly detected at high luminosity by both missions. This implies
an increase in flux of $\sim$2 orders of magnitude or more during this
$\sim$4 day window. We argue that this is likely due to a rapid rise in the
mass accretion rate, rather than to a transition from an extremely obscured
to an unobscured state. During the second epoch we observed the broadband
0.3--20.0\,\kev\ X-ray luminosity to be $(1.55 \pm 0.06) \times
10^{40}$\,\ergps, similar to the majority of the archival X-ray observations.
The broadband X-ray spectrum obtained from the second epoch is
inconsistent with the low/hard accretion state observed in Galactic black hole
binaries, but is well modeled with a simple accretion disk model incorporating
the effects of photon advection. This strongly suggests that, when bright,
\ngc\ is a high-Eddington accretor.
\end{abstract}


\section{Introduction}

Ultraluminous X-ray Sources (ULXs) are off-nuclear point sources with X-ray
luminosities that exceed the Eddington limit for the $\sim$10\,\msun\
`stellar-mass' black holes observed in Galactic black hole binaries (BHBs; \eg
\citealt{Orosz03}), \ie $L_{\rm X} > 10^{39}$\,\ergps. Multi-wavelength
observations have largely excluded highly anisotropic emission that could
artificially increase the estimated luminosity (\eg \citealt{Berghea10b, Moon11}).
The observed luminosities therefore require either the presence of larger black
holes than those observed in our own galaxy (\eg \citealt{Miller04,
Strohmayer09a, Zampieri09}), or  super-Eddington modes of accretion (\eg
\citealt{Poutanen07, Finke07}). For recent reviews focusing on ULXs, see
\cite{Roberts07rev} and \cite{Feng11rev}.

Although the majority of ULXs only have luminosities marginally in excess of
$10^{39}$\,\ergps\ (\citealt{WaltonULXcat, Swartz11}), and therefore likely
represent a high luminosity extension of the stellar mass BHB population
(\citealt{Middleton13nat, Liu13nat, Motch14nat}), a smaller population of
extreme sources have observed X-ray luminosities of $L_{\rm X} > 10^{40}$
\ergps\ (\eg \citealt{Farrell09, Walton4517, Jonker12}). The extreme
luminosities displayed by these sources are of substantial interest, and mean
they remain among the best candidates to host black holes more massive than
those observed in Galactic BHBs.

\ngc\ is a luminous member of this population of extreme ULXs. The source
was initially reported in the ULX catalogue presented in \cite{WaltonULXcat}
with a peak X-ray luminosity of $\sim$5$\times10^{40}$\,\ergps\ (see also
\citealt{Sutton12}). Although X-ray data on the edge-on spiral galaxy
NGC\,5907 is relatively sparse, since the discovery of ULX1 a number of
follow-up observations have been undertaken with \xmm, \chandra\ and
\swift, revealing the source to be variable by a factor of a few, confirming that
a single, accretion powered source dominates the observed X-ray flux 
(\citealt{Sutton13}). Based on data with a bandpass limited to $E<10$\,\kev,
the observed characteristics of \ngc\ below 10\,\kev\ appear to be broadly
consistent with a BHB in the sub-Eddington low/hard state (\citealt{Sutton12};
see \citealt{Remillard06rev} for details on the standard BHB accretion states),
implying the possible presence of a very massive black hole. However, the
highest quality soft X-ray data available tentatively suggest the presence of a
spectral break above $\sim$5\,\kev\ (\citealt{Sutton13}) which, if confirmed,
would be inconsistent with this accretion regime, and potentially identify
\ngc\ as a high-Eddington source (\citealt{Gladstone09}). 

The \textit{Nuclear Spectroscopy Telescope Array} (\nustar; \citealt{NUSTAR}),
in conjunction with \xmm, \suzaku\ and \chandra, has been providing the
first ever high quality broadband X-ray spectra for a sample of known ULXs
(see \citealt{Bachetti13, Bachetti14nat, Rana14, Walton13culx, Walton14hoIX}).
Here we present results from the recent \nustar\ and \xmm\ observations of
\ngc. The paper is structured as follows: section \ref{sec_red} describes our
observations and data reduction procedure, and sections \ref{sec_e1},
\ref{sec_e2} and \ref{sec_archive} describe the analysis performed. Finally, we
discuss our results and conclusions in section \ref{sec_dis}.

\section{Observations and Data Reduction}
\label{sec_red}

During 2013 the \nustar\ and \xmm\ X-ray observatories performed two
coordinated observations of \ngc, with some portion of the \nustar\ observation
simultaneous with \xmm\ in both cases (see Table \ref{tab_obs} for details).
Here, we outline our general data reduction procedure for these observations;
details specific to each epoch are given in sections \ref{sec_e1} and \ref{sec_e2}.

\subsection{NuSTAR}

\nustar\ performed two observations of \ngc\ in late 2013, referred to
throughout this work as epochs 1 and 2 (see Table \ref{tab_obs}; although the
first epoch is comprised of two OBSIDs, it is actually one continuous observation).
The start of the second observation is roughly a week after the start of
the first, however given the duration, the period between the end
of the first observation and the start of the second is only $\sim$4 days.
We reduced the \nustar\ data using the standard pipeline (\nupipeline),
part of the \nustar\ Data Analysis Software v1.3.1 (\nustardas; included in the 
\heasoft\ distribution as of version 6.14), and we use instrumental calibration
files from \nustar\ caldb v20131223 throughout. The unfiltered event files were
cleaned with the standard depth correction, significantly reducing the internal
background at high energies, and passages through the South Atlantic Anomaly
were removed. Source and background products were extracted from the cleaned
event files for both focal plane modules (FPMA and FPMB) using \nuproducts,
with the background primarily estimated from a blank area of the same
detector in each case (unless stated otherwise).

\subsection{XMM-Newton}

Each of the two \nustar\ observations was coordinated with a shorter
observation with \xmm, providing soft X-ray coverage down to $\sim$0.3\,\kev.
Data reduction was carried out with the \xmm\ Science Analysis System (\sas\
v13.5.0) in accordance with the standard prescription provided in the online
guide.\footnote{http://xmm.esac.esa.int/} The raw observation data files were
processed using \epchain\ and \emchain\ to produce cleaned event lists for
the \epicpn\ (\citealt{XMM_PN}) and \epicmos\ (\citealt{XMM_MOS}) detectors,
respectively. In this work, we use only single and double events (single to
quadruple events) for \epicpn\ (\epicmos), and exclude periods of high
background flares (adopting thresholds of 0.5 and 0.12\,ct s$^{-1}$ in
the 10--12\,\kev\ lightcurve from the full field-of-view for \epicpn\ and
each \epicmos\ detector, respectively). Science products were extracted using
\xmmselect, with the background estimated from areas of the same CCD free of
contaminating point sources. Redistribution matrices and auxiliary response files
were generated with \rmfgen\ and \arfgen. After performing the data reduction
separately for each of the two \epicmos\ detectors, and confirming their
consistency, these spectra were combined using the \ftool\ \addascaspec.

\begin{table}
  \caption{Details of the X-ray observations of NGC\,5907 ULX1 considered
in this work.}
\begin{center}
\begin{tabular}{c c c c c}
\hline
\hline
\\[-0.25cm]
Mission & OBSID & Start Date & Good Exposure\tmark[a] \\
& & & (ks) \\
\\[-0.3cm]
\hline
\hline
\\[-0.1cm]
\multicolumn{4}{c}{\textit{Epoch 1}} \\
\\[-0.2cm]
\nustar\ & 30002039002 & 2013-11-06 & 45 \\
\\[-0.225cm]
\nustar\ & 30002039003 & 2013-11-06 & 69 \\
\\[-0.225cm]
\xmm\ & 0724810201 & 2013-11-06 & 22/26 \\
\\[-0.1cm]
\multicolumn{4}{c}{\textit{Epoch 2}} \\
\\[-0.2cm]
\nustar\ & 30002039005 & 2013-11-12 & 113 \\
\\[-0.225cm]
\xmm\ & 0724810401 & 2013-11-12 & 24/34 \\
\\[-0.1cm]
\multicolumn{4}{c}{\textit{Archival}} \\
\\[-0.2cm]
\chandra\ & 12987 & 2012-02-11 & 11 \\
\\[-0.225cm]
\chandra\ & 14391 & 2012-02-11 & 12 \\
\\[-0.225cm]
\swift\ & 00032764001 & 2013-03-19 & 4 \\
\\[-0.225cm]
\swift\ & 00032764002 & 2013-04-03 & 4 \\
\\[-0.225cm]
\swift\ & 00032764003 & 2013-04-04 & 4 \\
\\[-0.225cm]
\swift\ & 00032764004 & 2013-04-06 & 3.5 \\
\\[-0.225cm]
\swift\ & 00032764005 & 2013-04-10 & 3.5 \\
\\[-0.225cm]
\swift\ & 00032764006 & 2013-05-04 & 4 \\
\\[-0.2cm]
\hline
\hline
\\[-0.15cm]
\end{tabular}
\\
$^{a}$ \xmm\ exposures are listed for the \epicpn/MOS detectors.
\vspace*{0.3cm}
\label{tab_obs}
\end{center}
\end{table}

\begin{figure*}
\hspace*{-0.25cm}
\epsscale{0.55}
\plotone{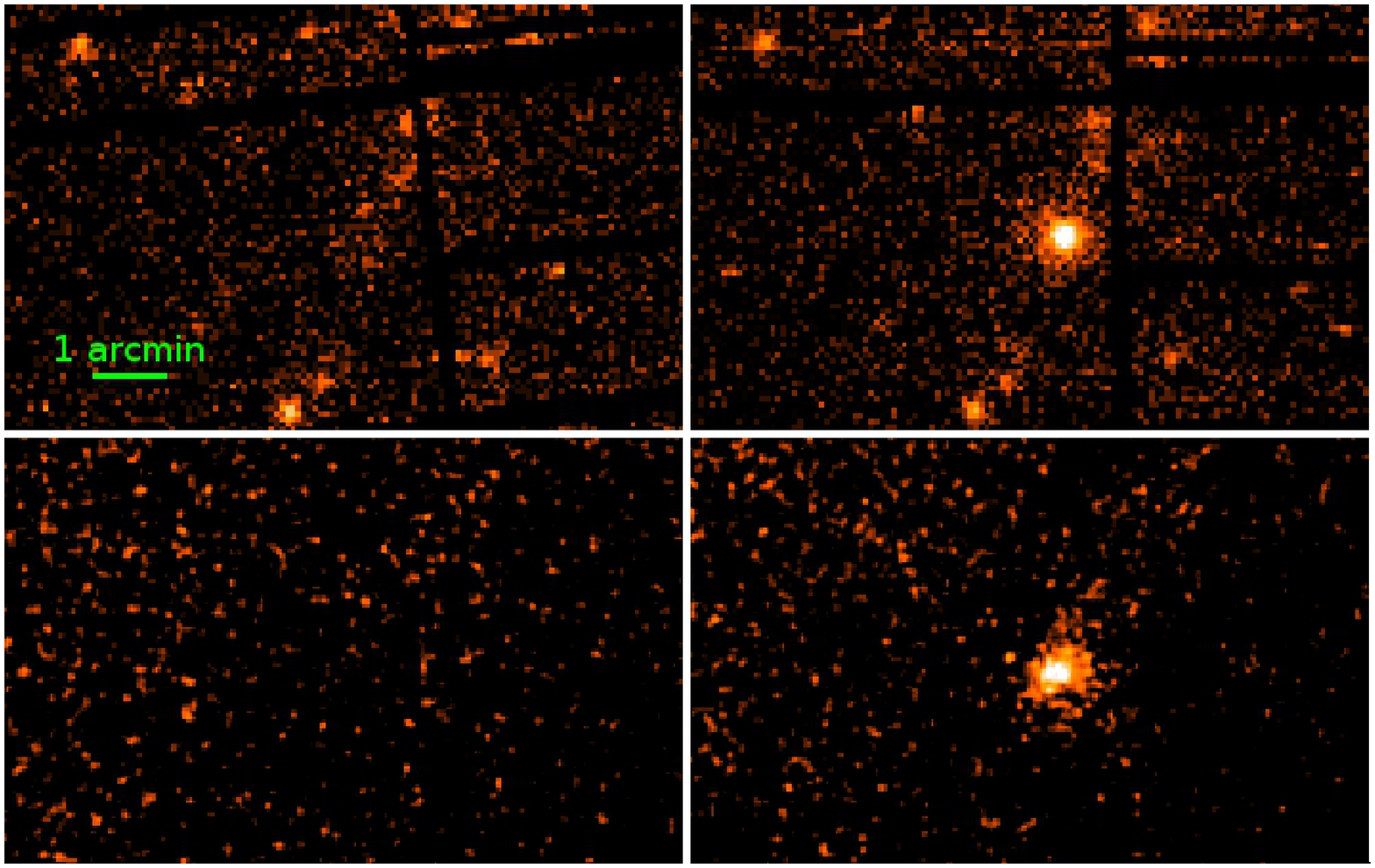}
\hspace*{0.75cm}
\plotone{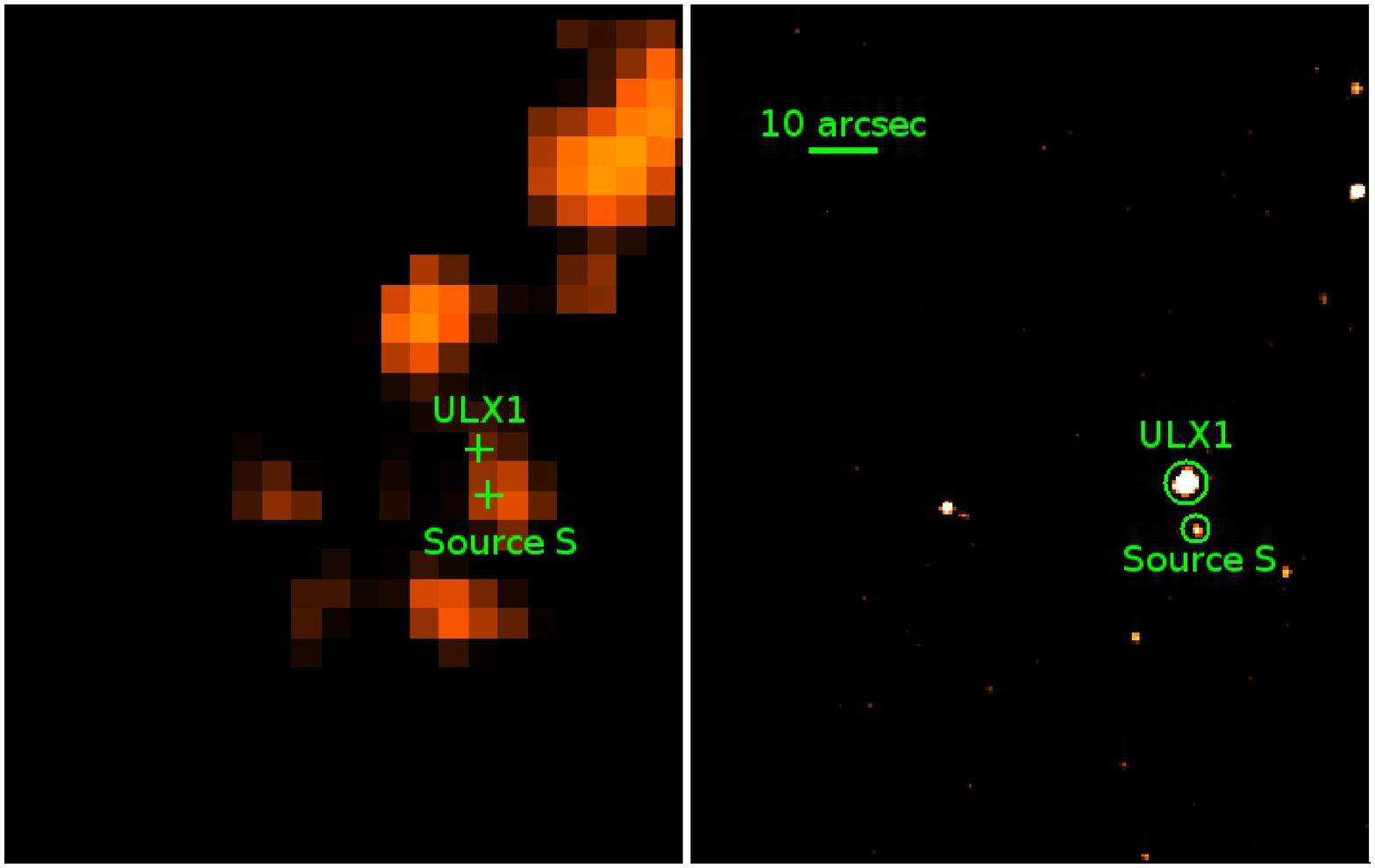}
\caption{
\textit{Left figure:} 4-panel image from the \xmm\ (\epicpn, \textit{top panels})
and \nustar\ (FPMA; \textit{bottom panels}) observations of \ngc\ from the first
(\textit{left panels}) and second epochs (\textit{right panels}). The ULX is clearly
detected in the second epoch, but is either very weak or absent in the first.
\textit{Right figure:} Zoom in on the immediate vicinity around the position of
the ULX in the \xmm\ image from the first epoch (\textit{left panel}), and the
same region from archival \chandra\ data (\textit{right panel}); these images
have been additionally smoothed for clarity. A number of faint sources are seen
in the \xmm\ data during this epoch, including one close to the position of the
ULX. However, this is likely to be dominatd by the faint source seen to the south
of the ULX in the \chandra\ image (\altsrc; see section \ref{sec_chan}).}
\vspace{0.2cm}
\label{fig_image}
\end{figure*}

Throughout this work we perform spectral analysis with \xspec\ v12.8.1
(\citealt{xspec}), and all models include neutral absorption from both the
Galactic column ($N_{\rm{H;Gal}}=1.21\times10^{20}$ cm$^{-2}$ towards
NGC\,5907; \citealt{NH}), and an intrinsic absorption column at the redshift of
NGC\,5907 ($z=0.002225$)\footnote{from the NASA Extragalactic Database:
http://ned.ipac.caltech.edu/} which is free to vary. Neutral absorption is
treated with \tbnew\footnote{\tbnewlink}, the latest version of the \tbabs\
absorption code (\citealt{tbabs}), with the appropriate solar abundances.
Uncertainties are quoted at the 90\% confidence level for one parameter of
interest, unless stated otherwise. Where possible, the \xmm\ and \nustar\ data
are modeled simultaneously with all physical parameters linked between the
different datasets; the spectral agreement between \xmm\ and \nustar\ in their
common 3--10\,\kev\ bandpass is known to be fairly good (\citealt{Walton13culx,
Walton14hoIX}), and we account for residual flux cross-calibration uncertainties
between the \epicpn\ and \epicmos\ detectors (\xmm) and FPMA and FPMB
(\nustar) by allowing multiplicative constants to float between them (fixing
\epicpn\ to unity).

\section{Epoch 1}
\label{sec_e1}

Based on the previously published observations, the X-ray emission from \ngc\
appears to be fairly persistent (\citealt{Sutton13}). However, during the first
epoch reported here, no single point source obviously dominates the emission
at the position of the ULX (Figure \ref{fig_image}, \textit{left panel}). Instead, the
soft X-ray emission observed by \xmm\ from the immediate vicinity of the ULX
appears to be comprised of a series of faint sources (Figure \ref{fig_image},
\textit{right panel}), which even in combination do not result in a detection with
\nustar. One of these sources is close to the known position of the ULX, and we
therefore extract the \xmm\ spectrum of this source from a circular region of
radius 12$''$ to avoid contamination from the other sources. Fewer than 50
net source counts are detected in total (\epicpn+\epicmos) in the
0.3--10.0\,\kev\ bandpass, so we rebin the data from this epoch to have a
minimum of 5 counts per bin (before background subtraction) to
maintain spectral coverage, and minimize the Cash statistic (\citealt{cstat}) when
analyzing these data. Here, the background contributes $\sim$45\% of
the total counts from the source region.

With so few counts it is not possible to undertake detailed spectral modeling,
but to estimate a source flux we model the spectrum with a simple absorbed
powerlaw, assuming $\Gamma=1.7$ based on previous \xmm\ observations
(\citealt{Sutton12}). We find an observed 0.3--10.0\,\kev\ flux of
$F_{\rm{Epoch1}}=11^{+5}_{-3}\times10^{-15}$\,\ergpcmsqps\ (throughout this
work fluxes are estimated with \cflux), substantially lower than any flux
observed from \ngc\ to date (\citealt{Sutton13}). For a distance to NGC\,5907
of 13.4\,Mpc (\citealt{Tully09}), the corresponding 0.3--10.0\,\kev\
luminosity is $2.4^{+1.1}_{-0.7}\times10^{38}$\,\ergps.

\begin{table}
  \caption{Best fit parameters obtained for the simple continuum models applied
  to the broadband spectrum observed from NGC\,5907 ULX1 during epoch 2.}
\begin{center}
\begin{tabular}{c c c c c c}
\hline
\hline
\\[-0.15cm]
Model: & \pl\ & \diskbb\ & \diskpbb \\
\\[-0.2cm]
\hline
\hline
\\[-0.1cm]
\nh\ ($10^{21}$\,cm$^{-2}$) & $9.3\pm0.6$ & $3.4\pm0.3$ & $6.7\pm0.7$ \\
\\[-0.225cm]
$\Gamma$ & $2.14\pm0.06$ & - & - \\
\\[-0.225cm]
$T_{\rm{in}}$ (\kev) & - & $1.90\pm0.07$ & $2.9\pm0.3$ \\
\\[-0.225cm]
$p$ & - & - & $0.55\pm0.02$ \\
\\[-0.2cm]
\hline
\\[-0.15cm]
$\chi^{2}$/DoF & 403/277 & 393/277 & 286/276 \\
\\[-0.2cm]
\hline
\hline
\\[-0.15cm]
\end{tabular}
\label{tab_results}
\end{center}
\end{table}

\section{Epoch 2}
\label{sec_e2}

We find that \ngc\ is very clearly detected by both \xmm\ and \nustar\ in the
second epoch (see Figure \ref{fig_image}), suggesting a remarkable flux
transition in a fairly short space of time (the end of the first \nustar\ observation
and the start of the second are separated by $\sim$4 days). We extracted spectra
using a circular region of radius 21$''$ for \xmm, chosen to simultaneously
maximize the signal-to-noise for the ULX and minimize the contamination from
the fainter sources nearby, and of radius 50$''$ for \nustar, given its larger PSF.
In order to improve the statistics at the highest energies, we combine the spectra
obtained by FPMA and FPMB with \addascaspec\ (after confirming their
consistency), and rebin all the spectra from this epoch to a minimum of 25 counts
per bin, minimizing \chisq\ in our analysis of these data. We obtain a detection in
\nustar\ up to just over 20\,\kev. 

\begin{figure*}
\hspace*{-0.5cm}
\epsscale{0.55}
\plotone{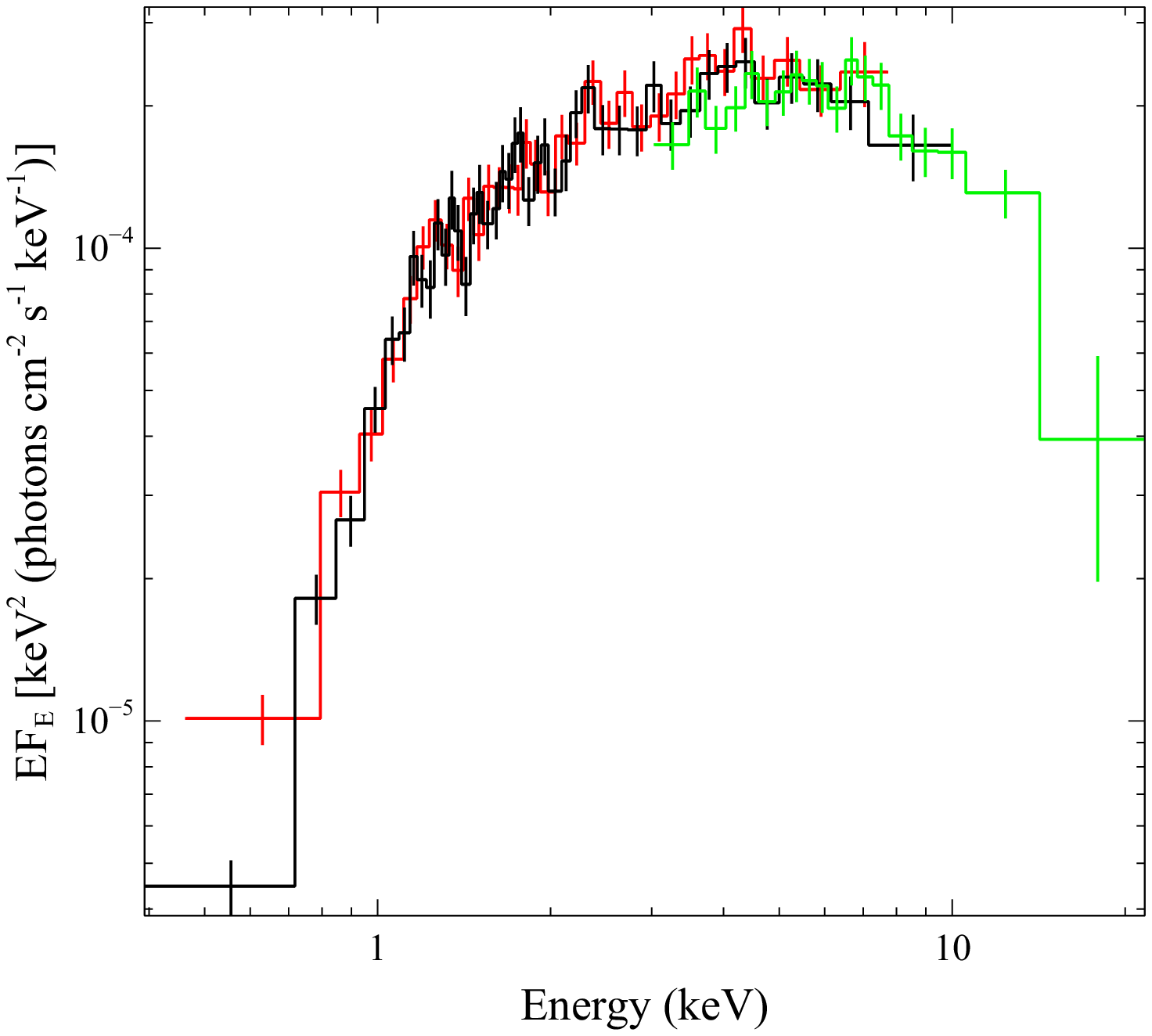}
\hspace*{0.75cm}
\plotone{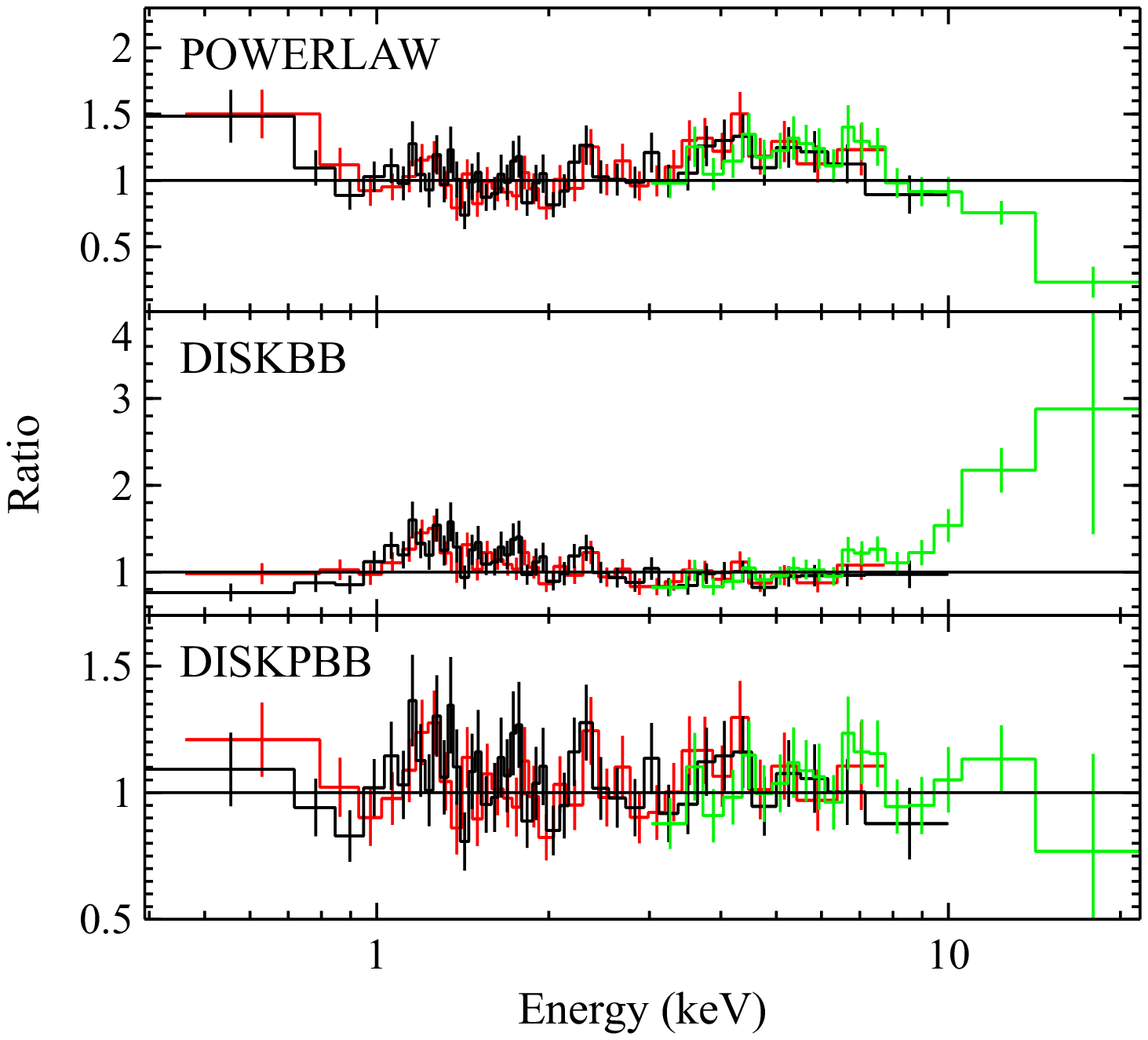}
\caption{
\textit{Left panel:} the broadband \xmm+\nustar\ X-ray spectrum of \ngc\
obtained from \etwo. The data have been unfolded through a model, which is
simply constant with energy. The \xmm\ \epicpn\ and \epicmos\ data are shown
in black and red, respectively, and the \nustar\ FPM data are shown in green.
\textit{Right panels:} data/model ratios for the continuum models applied to the
\etwo\ spectrum (see section \ref{sec_e2}). In all panels, the data have been
rebinned for visual clarity}
\label{fig_spec}
\end{figure*}

Figure \ref{fig_spec} shows the broadband 0.3--25.0\,\kev\ spectrum obtained
from \etwo. The \nustar\ data confirm the presence of a spectral break in the
$\sim$3--10\,\kev\ bandpass, similar to other ULXs observed by \nustar\ to
date (\citealt{Bachetti13, Rana14, Walton13culx, Walton14hoIX}). As the data for
\ngc\ from this epoch are of much lower quality compared to the broadband
observations of those other ULX targets, we limit our spectral analysis in this
work to simple continuum modeling of the time averaged spectrum; the
count rates obtained unfortunately do not permit detailed variability studies of
these data (0.12 and 0.04 ct s$^{-1}$ in 0.3--10.0\,\kev\ from \epicpn\ and
each \epicmos\ detector, and 0.008 ct s$^{-1}$ in 3--25\,\kev\ from each
\nustar\ FPM). Both Galactic and intrinsic neutral absorption are included in all
models. In addition, given the larger extraction region used for the \nustar\
data, we also investigate whether the results presented below are influenced by
undetected contamination from the other sources seen in the \eone\ \xmm\
data (Figure \ref{fig_image}), and we repeat our analysis using the \eone\
\nustar\ spectrum extracted from the \etwo\ source region as an alternative
background spectrum. We obtain consistent results with both background
estimations, and we therefore only present those obtained with the \etwo\
background for simplicity.

We first fit the broadband spectrum with an absorbed powerlaw model, but find
that this provides a poor fit (\rchi\ $=\chi^{2}/\rm{DoF} = 403/277$; see Table
\ref{tab_results}), and results in systematic residuals across the whole bandpass
(see Figure \ref{fig_spec}, \textit{right panel}). In particular, the model
overpredicts the highest energies, confirming the existence of spectral curvature
above $\sim$3\,\kev. We also find that a model consisting of a geometrically
thin, optically thick accretion disk (\diskbb, \citealt{diskbb}; see also 
\citealt{Shakura73}) provides a poor fit (\rchi\ $=393/277$), clearly
under-estimating the high energy flux. However, a slightly more complex
accretion disk model in which the radial temperature index ($p$) is free to vary
(\diskpbb, \citealt{diskpbb}) does provides a good fit across the full bandpass
(\rchi $=286/276$), and there is no obvious requirement for any additional 
spectral components. The radial temperature index of $p=0.55\pm0.02$
obtained (see Table \ref{tab_results}) is shallower than expected for a thin
accretion disk  ($p_{\rm{thin}}=0.75$), consistent with an accretion disk in which 
strong advection of radiation occurs, as may be expected at very high accretion
rates where radiation pressure should dominate and modify the disk structure
(\eg \citealt{Abram88}).

Based on the \diskpbb\ model we find an observed 0.3--20.0\,\kev\ X-ray flux
for \ngc\ during the second epoch of $F_{\rm{Epoch2}} = (7.2 \pm 0.3) \times
10^{-13}$\,\ergpcmsqps, $\sim$87\% of which falls in the 0.3--10.0\,\kev\
bandpass. The observed fluxes of the various datasets considered in this work
are summarized in Table \ref{tab_flux}. This equates to a broadband X-ray
luminosity of $(1.55 \pm 0.06) \times 10^{40}$\,\ergps\ even before any
absorption correction. While extreme for the ULX population, this is still a factor
of $\sim$2--3 lower than the peak luminosity of $\sim$5 $\times
10^{40}$\,\ergps\ (0.3--10.0\,\kev) that has been observed from this source
(\citealt{Sutton13}).

\section{Archival Data}
\label{sec_archive}

\subsection{Chandra Imaging and Astrometry}
\label{sec_chan}

To determine the ULX flux during \eone\ it is essential to address the level of
source confusion. We inspected the archival data obtained with the \chandra\
observatory (\citealt{Chandra}; see Table \ref{tab_obs}), which reveal a faint
source $\sim$6--7$"$ to the South of \ngc\ (hereafter \altsrc; see Figure
\ref{fig_image}). In order to assess whether the \xmm\ detection could be
associated with \altsrc, rather than with the ULX, we extracted the \chandra\
spectrum of \altsrc. Both \chandra\ observations were taken in the Timed Event
mode, and we extracted spectra from the ACIS-S detector
(\citealt{CHANDRA_ACIS}) using the standard pipeline in CIAO v4.6. The source
spectrum was obtained from a circular region of radius $\sim$2$''$, while the
background was extracted from a larger circular region of radius $\sim$13$''$
that was free from any other contaminating sources. The ACIS spectra from the
two observations were combined using \addascaspec. Very few net
source counts are detected (fewer than 15), so we rebin the spectrum to a
minimum of 2 counts per bin and again minimize the Cash statistic when
considering these data. In this case, the background contributes only
$\sim$10\% of the total counts from the source region. Applying the same
model used for the detection from \eone\ to the \chandra\ data for \altsrc, we
find an observed 0.3--10.0\,\kev\ flux of $F_{\rm{\altsrc}} = (7\pm 3) \times
10^{-15}$ \ergpcmsqps. This is consistent with that obtained for the \eone\
\xmm\ observation (section \ref{sec_e1}), which may suggest a common origin.
Were this to be the case, the variability displayed by \ngc\ between epochs 1 and
2 would be even greater than implied by the measured \xmm\ fluxes (sections
\ref{sec_e1} and \ref{sec_e2}).

Unfortunately we were not able to reliably correct the \xmm\ astrometry
in \eone\ against \chandra\ directly, owing to a lack of sources within the
overlapping \xmm\ and \chandra\ sky coverage. However, there are just about
sufficient sources in the \xmm\ field-of-view to determine whether there is any
astrometric offset between epochs 1 and 2. To this end, we generated source lists
for the \epicpn\ detector from both epochs using {\small EDETECT\_CHAIN}, and
then computed the astrometric correction between the two epochs using {\small
EPOSCORR} (both part of the \xmm\ \sas). This found 5 robust source matches
(note that the ULX itself was excluded from the matching procedure). The the
best solution to the source matching found the offset between epochs to be
$(2.7 \pm 0.1)''$ in Right Ascension and $(1.9 \pm 0.1)''$ in Declination, with no
rotation component, such that \etwo\ is shifted to the South and the  West
relative to \eone. We used these offsets to correct the position obtained for ULX1
with {\small EDETECT\_CHAIN} in \etwo, when ULX1 dominated the observed
emission, to the coordinate system registered to the image from \eone.
Having done so, we then also worked out the position of \altsrc\ in this image,
using the relative positions of ULX1 and \altsrc\ determined from the \chandra\
data. These corrected positions are shown in the \xmm\ image of \eone\ in
Figure \ref{fig_image}. The peak of the faint emission from \eone\ is more
consistent with the position of \altsrc, although there may additionally be some
even fainter extent to the emission towards the position of the ULX. Therefore,
while there may still be some contribution from ULX1, we conclude that the
emission detected in \eone\ is likely dominated by \altsrc.

\begin{table}
  \caption{Observed fluxes from the datasets analyzed in this work. The {\it 
Chandra} flux for Source S is also listed for comparison.}
\begin{center}
\begin{tabular}{c c c}
\hline
\hline
\\[-0.2cm]
Dataset &  0.3--10.0\,\kev\ Flux & 0.3--20.0\,\kev\ Flux \\
\\[-0.25cm]
& \multicolumn{2}{c}{($10^{-15}$\,\ergpcmsqps)} \\
\\[-0.3cm]
\hline
\hline
\\[-0.2cm]
Epoch 1\tmark[a] & $11^{+5}_{-3}$ & - \\
\\[-0.2cm]
Epoch 2 & $630^{+30}_{-20}$ & $720 \pm 30$ \\
\\[-0.2cm]
2013 \swift\ & $30^{+20}_{-10}$ & - \\
\\[-0.2cm]
\hline
\\[-0.2cm]
\altsrc\ (\chandra) & $7 \pm 3$ & - \\
\\[-0.2cm]
\hline
\hline
\\[-0.2cm]
\multicolumn{3}{l}{$^{a}$ Likely dominated by emission from \altsrc, see section
\ref{sec_chan}.}
\end{tabular}
\label{tab_flux}
\end{center}
\end{table}

\begin{figure*}
\hspace*{-0.5cm}
\epsscale{1.15}
\plotone{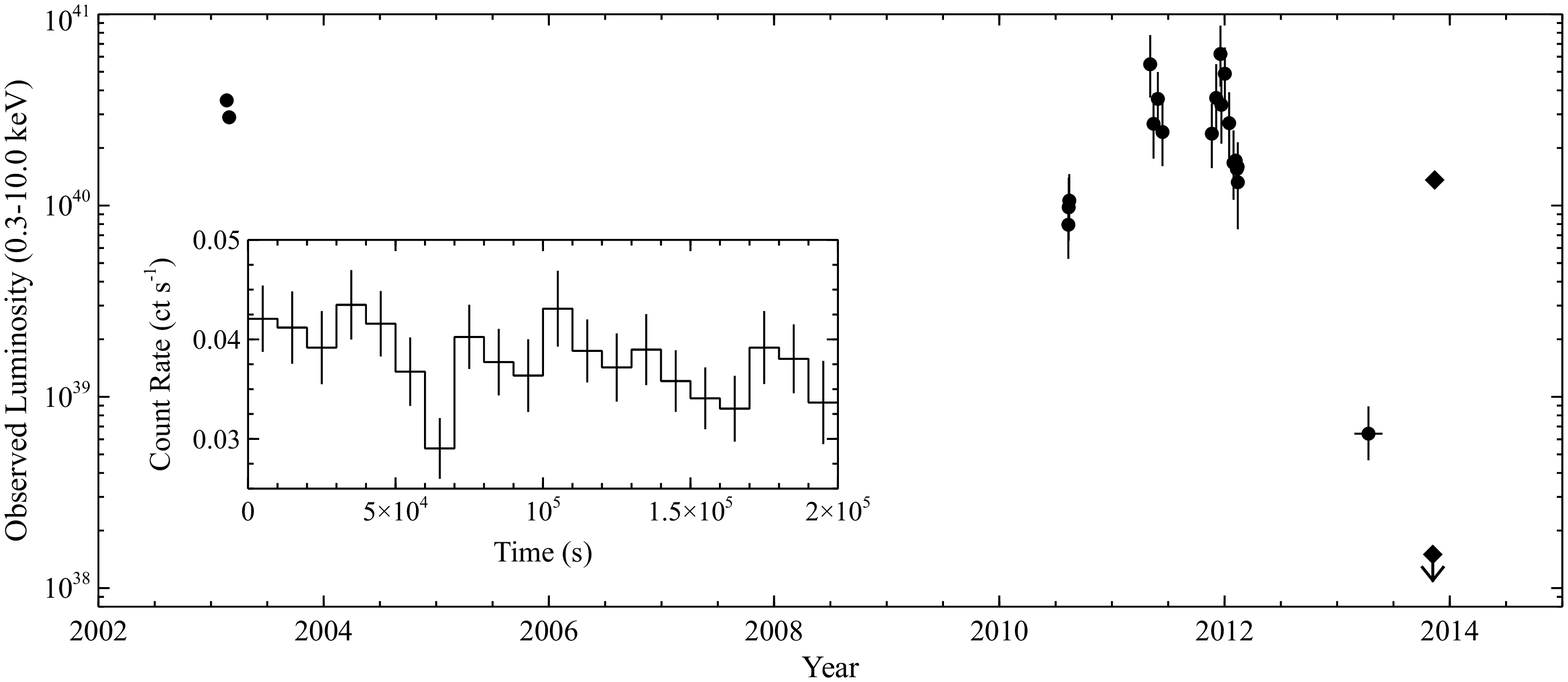}
\caption{
The long-term lightcurve observed from \ngc\ since its discovery by \xmm\ in
2003 (adapted from \citealt{Sutton13}). Our most recent \xmm\ and \nustar\
observations are indicated with diamonds, while archival observations with
\xmm, \chandra\ and \swift\ are indicated with circles (see \citealt{Sutton13}
for details). For consistency with \cite{Sutton13}, we show 1$\sigma$ errors
here. Given that the detected emission from \eone\ is dominated by
\altsrc, we adopt a conservative upper-limit on the flux from ULX1 during this
epoch of half of the upper bound on the total flux observed by \xmm, but we
assume the combined 2013 \swift\ data does represent a detection of \ngc.
\textit{Inset}: the \nustar\ (FPMA+FPMB) lightcurve observed from \ngc\ during
\etwo\ (10\,ks bins), displaying a relatively stable flux throughout the
observation.}
\label{fig_lc}
\end{figure*}

\subsection{2013 Swift Snapshots}

A series of six short observations were undertaken in 2013 by \swift\
(\citealt{SWIFT}), prior to our broadband observations (between March and May,
see Table \ref{tab_obs} for details) which were not included in \cite{Sutton13}.
We inspected these data in order to provide further context for the extreme low
flux observed in \eone. Although the earlier 1-2\,ks \swift\ snapshots presented
in \cite{Sutton13} easily detected \ngc, it is not obviously detected by the XRT in
any of the individual 2013 observations, despite their longer exposure (typically
$\sim$4\,ks), indicating a prolonged period at low flux earlier in 2013. However,
although individual observations do not provide a clear detection, if we stack the
2013 observations, we do find a clear detection of a source at the position of
\ngc\ in the combined dataset. 

We extracted the spectrum from this stacked dataset, following the standard XRT
reduction
guide\footnote{http://swift.gsfc.nasa.gov/analysis/xrt\_swguide\_v1\_2.pdf} in
order to estimate the average source flux during this period. All the observations
were obtained in the standard photon-counting mode, and we extracted the
source spectrum from the same region used for the \eone\ \xmm\ data,
estimating the background from a much larger region of radius 180$''$ avoiding
the plane of NGC\,5907 and other contaminating point sources. We used the
latest XRT redistribution matrix available in the \swift\ CALDB, and generated the
anciliary response file as standard for a point source on axis with \xrtmkarf,
correcting for PSF losses to account for the small extraction region. Very
few net source counts are detected ($\sim$10), so we rebin to a minimum of 2
counts per bin, and minimize the Cash statistic when analyzing these data.
Here the background contribution is again very small, only $\sim$10\%
of the total counts from the source region.

Applying the same model used for the \xmm\ detection from \eone, we
find an average 0.3--10.0\,\kev\ flux of $F_{\rm{Swift}} = 30^{+20}_{-10}
\times 10^{-15}$\,\ergpcmsqps\ during the 2013 observations, implying a
luminosity of $6^{+4}_{-3} \times 10^{38}$\,\ergps. The \swift\ detection is a
factor of $\sim$3 brighter than both the \eone\ \xmm\ detection, and \altsrc\
during the \chandra\ observations (albeit with admittedly large uncertainties).
Although we obviously cannot exclude the possibility that \altsrc\ is also
variable, given the extreme level of variability \ngc\ is now known to exhibit it
seems natural to assume the variability between the 2013 \swift\ data and
\eone\ is also driven by \ngc. This would suggest the \swift\ data constitute a
detection of \ngc, even if the \eone\ \xmm\ data potentially do not.

\section{Discussion and Conclusions}
\label{sec_dis}

We have presented an analysis of two epochs of broadband X-ray observations
of the extreme ULX \ngc\ undertaken with \nustar\ and \xmm. These
observations reveal an astonishing level of X-ray flux variability between the two
epochs. Although the source is not detected by \nustar\ in the first epoch, a
clear detection is obtained in the second, providing the first constraints on the
hard X-ray ($E>10$\,\kev) emission. Broadly similar to the results obtained for
other ULXs observed by \nustar\ to date (\citealt{Bachetti13, Rana14,
Walton13culx, Walton14hoIX}, Mukherjee \etal, \textit{in prep.}), we find the hard
X-ray emission from \ngc\ to be very weak relative to that at lower energies (see
Figure \ref{fig_spec}). During this epoch the broadband spectrum is not consistent
with a $\sim$10$^{3-4}$\,\msun\ intermediate mass black hole (IMBH) accreting
in the low/hard state displayed by Galactic BHBs. Instead, the spectrum is well
modeled with an advection dominated accretion disk (\eg \citealt{Abram88}),
confirming the spectral cutoff tentatively suggested by the archival \xmm\ data
(\citealt{Sutton13}), and implying that \ngc\ may be accreting at a very high,
possibly super-Eddington rate.

The most intriguing aspect of these observations is the extreme level of flux
variability. We show in Figure \ref{fig_lc} a long-term X-ray lightcurve for \ngc,
adapted from \cite{Sutton13} to include our \xmm\ and \nustar\ observations
(and also the additional 2013 \swift\ data). Even assuming the faint source
detected by \xmm\ in \eone\ is \ngc, its observed flux varied by $\sim$2
\ordmag\ in $\sim$4 days (the flux appears stable throughout \etwo, see the
inset in Figure \ref{fig_lc}). However, given the flux observed by
\chandra\ and the astrometric offset between the \eone\ and \etwo\ \xmm\
observations, the emission detected in \eone\ is likely dominated by \altsrc\
(see section \ref{sec_chan}), implying an even more extreme variation from
ULX1. While there are transient ULXs that show orders of magnitude of
variation (similar to low mass X-ray binaries, \eg\ \citealt{Middleton13nat}),
even at peak luminosity these tend to be the fainter members of the ULX
population. The brighter ULXs tend to be variable only by a factor of $\sim$a
few, broadly similar to high mass X-ray binaries, (\eg\ \citealt{Kaaret09m82,
Kaaret09ulx, Kong10, Walton13culx}). Until this work, the behavior observed
from \ngc\ was similar to this latter population, consistent with its extreme
luminosity.

Given the relatively persistent behavior observed previously one tantalizing
possibility for the extreme low state observed in \eone\ is that it may represent
an eclipse of the X-ray source, perhaps by a stellar companion or by a warped
outer accretion disk. If this were the case, the lack of a hard X-ray detection with
\nustar\ suggests the medium eclipsing \ngc\ is extremely thick
($N_{\rm{H}}\sim10^{24}$\,cm$^{-2}$ or more). However, the duration of the
\nustar\ observation is $\sim$2 days (the low-earth orbit results in a
$\sim$50\% observing efficiency; \citealt{NUSTAR}), and we do not know how
long this low flux persisted before the \nustar\ observation, thus if this were an
eclipse by the companion star the orbital period would have to be $\gg$2 days.
For comparison, the orbital period of the eclipsing black hole binary IC\,10 X-1,
the most massive dynamically constrained stellar remnant measured to date, is
$\sim$34 hours (\citealt{Prestwich07, Silverman08}), and the X-ray eclipses last
$\sim$30\% of this period, significantly shorter than the \nustar\ observation.
The other eclipsing BHB system known, M\,33 X-7, has an orbital period of
$\sim$3.5 days, with the X-ray eclipses lasting $\sim$15\% of this
(\citealt{Pietsch06, Orosz07}). Nevertheless, GRS\,1915+105, one of the Galactic
sources widely considered most analogous to ULXs, has a much longer orbital
period ($\sim$34 days, \citealt{Steeghs13}), and several works have suggested
that ULXs might plausibly have very long orbital periods (up to $\sim$100 days
or more; \citealt{Podsiadlowski03, Pooley05, Madhusudhan08}) if they accrete
from evolved stellar companions via Roche-lobe overflow, as suggested 
by the lack of iron emission (\citealt{Walton12ulxFeK, Walton13hoIXfeK}). If the
eclipse is by some other material, e.g. the outer regions of a warped accretion
disk rather than the stellar companion, this may precess on long, super-orbital
timescales, and would ease  the requirement for a long orbit. A scenario along
these lines was proposed as a potential explanation for the rare (and generally
less extreme) dips observed from the well studied, soft-spectrum ULX
NGC\,5408 X-1 (lasting up to a few days, \citealt{Grise13}). This behavior is
perhaps analogous to the `dipping' low-mass X-ray binaries (\eg
\citealt{DiazTrigo09}).

The \swift\ observations earlier in 2013 represent an additional period of low
flux that appears to last $\sim$6 weeks (although we note the infrequent
observing cadence during this period). However, the flux during \eone\ is
fainter than that from the combination of these \swift\ observations though,
and likely substantially so given that the \eone\ emission is dominated
by \altsrc. It may therefore be possible that we are observing the effect of an
eclipse imprinted on top of a strong level of intrinsic variability. While it may
not be possible to conclusively rule out an explanation along these lines with
the current data, we do not consider this scenario to be particularly likely.
Nevertheless, it is interesting to note that if this were the case, it would imply
that we are viewing \ngc\ at a high inclination. This is at odds with the
suggested framework of the wind dominated `ultraluminous state' for
super-Eddington accretion in which hard spectrum ULXs similar to \ngc\ are
viewed close to face on, such that the hot inner regions of the accretion 
flow are visible. For edge-on sources, the inner regions would instead be
obscured by cooler material in a large scale-height wind launched from the
disk, resulting in a soft X-ray spectrum (\citealt{Sutton13uls}; Middleton et al.
2014, \textit{submitted}), inconsistent with that observed. If the variability
does result from a high inclination, this would thus favor the `patchy disk'
scenario suggested by \cite{Miller14}.

If the extreme rise in flux is not related to the end of an obscuration event, it
must instead be caused by a rapid increase in the mass accretion rate onto \ngc.
Such rapid increases in flux are occasionally seen in Galactic X-ray binaries, and
appear in some instances to be related to increased accretion, rather than
obscuration (\eg\ XTE\,J1701-407: \citealt{Degenaar11atel, Pawar13},
4U\,1700-377: \citealt{Smith12}). In the most extreme cases, super-fast X-ray
transients (SFXTs) can flare by orders of magnitude in an extremely short period
of time (hours or days, \eg \citealt{Sidoli09}), although these are very
short-lived flare events at much lower luminosities which differ markedly from
the behavior of \ngc.

One scenario that can result in rapid rises in accretion rate is for the
system to have a highly eccentric orbit, with large accretion bursts triggered
close to periastron, broadly similar to Be/X-ray binaries (\eg \citealt{Reig11,
Casares14}). This scenario is currently the leading interpretation for the almost
periodic outbursts exhibited by the most extreme ULX observed to date,
ESO\,243-49 HLX1 ($L_{\rm X,peak} \sim 10^{42}$\,\ergps), which displays a
repeated `fast-rise-exponential-decay' (FRED) outburst profile with similarly
large and rapid flux increases to those seen in our observations of \ngc\ (\eg
\citealt{Lasota11, Webb14rev}, although see \citealt{King14hlx}). Given the poor
sampling to date of the long-term behavior of \ngc, it is possible that
something similar could be occurring in this case, albeit at a lower absolute
luminosity, and with the source rising into a potentially super-Eddington state
(in contrast, HLX-1 seems to exhibit the standard evolution shown by
sub-Eddington Galactic binaries during its outbursts; \citealt{Servillat11,
Godet12}). Thus it may be possible for the extreme variability observed to be
related to the orbit of the system, without being associated with an eclipse event.
In light of these observations, extended monitoring of this remarkable source is
strongly recommended in order to test this exciting possibility.

\section*{ACKNOWLEDGEMENTS}

The authors would like to thank the reviewer for their positive and useful
feedback, which helped improve the manuscript, as well as Diego Altamarino for
useful discussions. MB and DB acknowledge financial support from the French
Space Agency (CNES). This research has made use of data obtained with the
\nustar\ mission, a project led by the California Institute of Technology (Caltech),
managed by the Jet Propulsion Laboratory (JPL) and funded by NASA, and has
utilized the \nustar\ Data Analysis Software (\nustardas) jointly developed by the
ASI Science Data Center (ASDC, Italy) and Caltech (USA). This research has also
made use of data obtained with \xmm, an ESA science mission with instruments
and contributions directly funded by ESA Member States and NASA, and NASA's
\chandra\ and \swift\ satellites. 

{\it Facilites:} \facility{NuSTAR}, \facility{XMM}, \facility{Chandra}, \facility{Swift}

\bibliographystyle{/Users/dwalton/papers/mnras}

\bibliography{/Users/dwalton/papers/references}

\label{lastpage}

\end{document}